# DistriFS: A Platform and User Agnostic Approach to File Distribution

**Julian Boesch**
jibtech1999@gmail.com

**ABSTRACT**

In an age where the distribution of information is crucial, current file sharing solutions suffer significant deficiencies. Popular systems such as Google Drive, torrenting and IPFS suffer issues with compatibility, accessibility and censorship. This paper introduces DistriFS, a novel decentralized approach tailored for efficient and large-scale distribution of files. The architecture of DistriFS is grounded in three foundational pillars: scalability, security, and seamless integration. The proposed server implementation harnesses the power of Golang, ensuring near-universal interoperability across operating systems and hardware. Moreover, the use of the HTTP protocol eliminates the need for additional software to access the network, ensuring compatibility across all major operating systems and facilitating effortless downloads. The design and efficacy of DistriFS represent a significant advancement in the realm of file distribution systems, offering a scalable and secure alternative to current centralized and decentralized models.

**KEYWORDS:** distributed computing, file distribution, decentralization

# 1 INTRODUCTION

In the past decade, the need for a file distribution solution that can handle millions of downloads while circumventing state censorship has become more evident than ever. The fundamental problem addressed by DistriFS is the limited efficiency and accessibility of current file sharing systems. File sharing in the digital era is essential, yet it is hindered by protocol restrictions, geographic censorship, and limitations imposed by user operating systems. The issues with most file sharing services today lie in the inherent design of these systems, which often rely on centralized architectures. These centralized systems, while easy to create and use, incur significant maintenance costs and pose risks of single points of failure (Johnson et al., 2008). Users seeking alternatives may turn to decentralized and self-hosted solutions to mitigate many of the shortcomings of centralized architectures. However, these systems are vulnerable to a different set of issues, such as major limitations in their ease of use and accessibility.

# 2 CURRENT WORK

Currently, centralized platforms such as Google Drive and Dropbox occupy most of the market share of file distribution technologies, as they have the lowest barrier to entry (most of these platforms offer free tiers) and intuitive, user-friendly interfaces. However, this ease of use comes at a cost: the companies that provide these services are extremely vulnerable to control and takedown by governmental bodies, especially in countries with stringent internet regulations like China, Russia, and Iran. Google publicizes takedown requests in transparency reports - looking at the statistics they provide, Russia alone has sent over 100,000 takedown requests to Google in the past decade (Brown, 2022). In addition to the risk of takedown, centralized file sharing solutions also have strict quotas. For example, Google Drive and Dropbox impose limits on uploading, copying, sharing and downloading files. The users who share and store the most are subject to the harshest limitations and policy changes, oftentimes with little or no notice - even paid customers, businesses and institutions (Amadeo, 2023).

Many privacy-conscious users may turn to existing decentralized solutions, such as torrenting and the InterPlanetary FileSystem (IPFS). While such alternatives to centralized file-sharing services are in use, their implementation often falls short of the user-friendly experience offered by centralized counterparts. Additionally, torrenting is often impeded by firewall restrictions, due to an assumption made by many governments and ISPs that torrenting traffic is solely for downloading illegal content. The majority of users are deterred from using torrenting due to these barriers, often resorting to paid VPNs or abandoning the method entirely (Morris, 2009). Other more recent solutions, such as IPFS, circumvent firewalls and government censorship more effectively. However, they lack ease-of-use and accessibility, and demonstrate inefficiencies in downloading less popular files (Benet, 2014). Shortcomings in accessibility exclude a minority of users, particularly those with disabilities and those using non-mainstream operating systems and hardware (Burda et al., 2013). While IPFS employs a similar architecture to DistriFS with the use of browser-based downloads, it introduces additional complexity by needing to translate these HTTP requests into its native TCP protocol[1]. This translation often results in downtime and timeout issues, particularly under heavy traffic conditions (Wan et al., 2017). In the academic sphere, studies like "Frangipani" (Thekkath, 1997) have delved into decentralized file systems, examining their potential and limitations. However, these studies have not fully addressed the specific challenges of creating a practical system that is both user-friendly and privacy-respecting, a key focus of DistriFS.

The most technically inclined of users may lean towards self-hosted platforms such as OwnCloud, NextCloud and Seafile to overcome the limitations and risk of trusting closed-source platforms. However, these alternatives are still centralized, and thus are vulnerable to a different set of data loss issues. Self-hosted platforms are vulnerable to physical drive failure, natural disasters and ransomware. While users are recommended to take mitigation steps like keeping regular backups and monitoring hard-drive health, very few individuals can afford ISO business continuity certifications[2] and professional audits to verify the security of their systems.

---

[1] DC++, a P2P file-sharing client which has many of the attributes of DistriFS, has the same complexity problem. Additionally, raw TCP protocols make it difficult to create a native web port of the protocol.

# 3 METHODS & DESIGN OF THE SYSTEM

The proposed solution is a distributed file system that integrates the benefits of decentralized sharing with an easy-to-use interface, prioritizing security and universal accessibility. Unlike torrenting, DistriFS operates over HTTP, offering a more robust, universally accessible protocol. The project can circumvent censorship on both the part of companies and governments by allowing and encouraging users to host their own indexers. Indexers are applications coded in Golang that keep track of files and their corresponding servers, offering readily available download links. Indexers accept basic JSON requests through any HTTP/S client. While the client has the option of being completely web-based, it is not possible for the indexer and server to run fully on the web due to restrictions of the browser sandbox. Despite this restriction, my choice of coding language creates the maximum compatibility possible with a vast spectrum of systems. Golang, developed by Google, is a language built upon strict compatibility standards and the principle that programs should be able to compile to a single executable file with all dependencies bundled (one executable file is provided for each operating system and architecture). This feature allows it to run on all popular platforms[3], allowing for plug-and-play operation of the indexer and server, even on obscure system setups (Cox et al., 2022).

DistriFS differentiates itself from existing solutions by ensuring that files are cryptographically verified by both the indexer and the client using the SHA256 hashing algorithm. Clients can easily verify the hashes of the files they download using an integrated feature in the client, allowing malicious routes (e.g., both the server and the indexer are compromised) to be identified and blocked. Finally, an integration with ClamAV, an open-source

---

[2] Such as ISO 22301:2019 certifications obtained by Google Cloud and Dropbox
[3] Namely: Android, Windows, iOS, Linux, macOS and FreeBSD. Golang supports 23 different architectures, including processors used by edge cases such as microcontrollers and supercomputers

antivirus, offers a free and privacy-conscious approach to virus detection. This approach to security provides both proactive and reactive solutions to malware.

By employing a "sensible defaults" configuration, DistriFS strikes a balance between user-friendliness and decentralization. This aspect is crucial, as it makes the system accessible to less tech-savvy users while providing options for those who prioritize privacy and security. Options such as DistriFS-over-Tor and multi-indexer searches (to verify hashes when searching names of files) are not needed by most users, but will be desired for the most privacy-conscious for fully anonymous and secure downloading. Additionally, advanced security settings would be desired should it be used in a government or corporate setting. The configurable and modular nature of DistriFS allows for such security settings to be configured without modifying the source code. Moreover, the system's design allows for a no-client architecture: users do not have to install any software to download files from the network, allowing DistriFS to have the same ease of use as simple HTTP servers, with added decentralization and security. This approach ensures that DistriFS can be utilized by users with varied technical proficiencies and resources[4].

## 3.1 DEPLOYMENT AND TESTING

The existing code will be made available on GitHub for the general public to view and suggest changes (Dabbish et al., 2012). This will allow people to report security issues and bugs, and suggest code changes. Additionally, all code history is tracked on GitHub, allowing people to track my changes across versions. GitHub Actions, a Continuous Integration (CI) server, will be used to certify that the compiled code has not been tampered with, allowing for the user to independently verify that DistriFS releases are safe to run. This robust system allows code changes and downloads to be tracked in an organized and secure manner.

After refining the prototype codebase into a more refined, readable and semantically correct form, rigorous code review and testing will be required. To make sure that each new version of DistriFS is ready for release, I will use multiple testing strategies. Unit tests (tests on a singular component of the system) will be used to ensure the usability of each system through a set of robust tests on every API route. Integration tests (tests to make sure that all components of the system are working together as expected) will serve as an additional check to make sure that the server is operational and performing up to expectations for download speed, ping and reliability. Additionally, load tests will be employed to evaluate how many clients and requests the server can handle (Munjal et al., 2015).

---

[4] Due to browser limitations, users downloading through a direct link will have to verify file hashes themselves. The official web and desktop clients do this automatically.

Finally, the deployment of multiple example servers will be necessary to showcase the system's capabilities. These servers will host diverse content types (such as software and ISO files)[5], providing a real-world testing ground for performance. Key performance indicators such as download speed, server uptime, and resource usage will be monitored and analyzed, offering tangible evidence of the system's efficiency. To validate the system's technical feasibility, comparative benchmarks will be established against existing file-sharing systems. These benchmarks will involve conducting tests to compare download speeds, resource utilization, and scalability. Additionally, the ability to scale will also be considered, as systems such as traditional web servers are more difficult to scale than distributed solutions. Replicable example code will be provided as evidence of test fairness and methodology.

### 3.2    INDEXER ARCHITECTURE

The indexer is one of the key components of a distributed and secure filesystem. It will be written to handle billions of files, and thousands of concurrent connections. The database used in this project is BadgerDB[6], based on the WISCKEY architecture (Lu et al., 2017). BadgerDB uses Log-Structured Merge (LSM) trees, which leverage the high sequential write speeds of SSDs to optimize specifically for newer hardware. The LSM-tree architecture stores all writes to the database in memory, and writes data periodically and sequentially to a database file. By grouping writes into "levels" of fixed size, the database is able to cache frequently used values.

---

[5] Such content types are needed to test the network's performance when downloading thousands of small files
[6] BadgerDB also is natively written in Golang, and works without any external software. This makes it an ideal alternative to SQLite (hard to scale, uses complex queries) and MongoDB (requires external software)

An optimized caching algorithm allows the server to respond quickly to the most queried data that is stored in the database - metadata and hashes. Files themselves are not stored in the database, so as to not duplicate stored data. Additionally, this approach allows files to instantly appear on the client, without any indexing or other performance-costly operation (Jain et al., 2017).

## 3.3 CLIENT ARCHITECTURE

**Fig 1. DistriFS Indexer-Client Relationship**

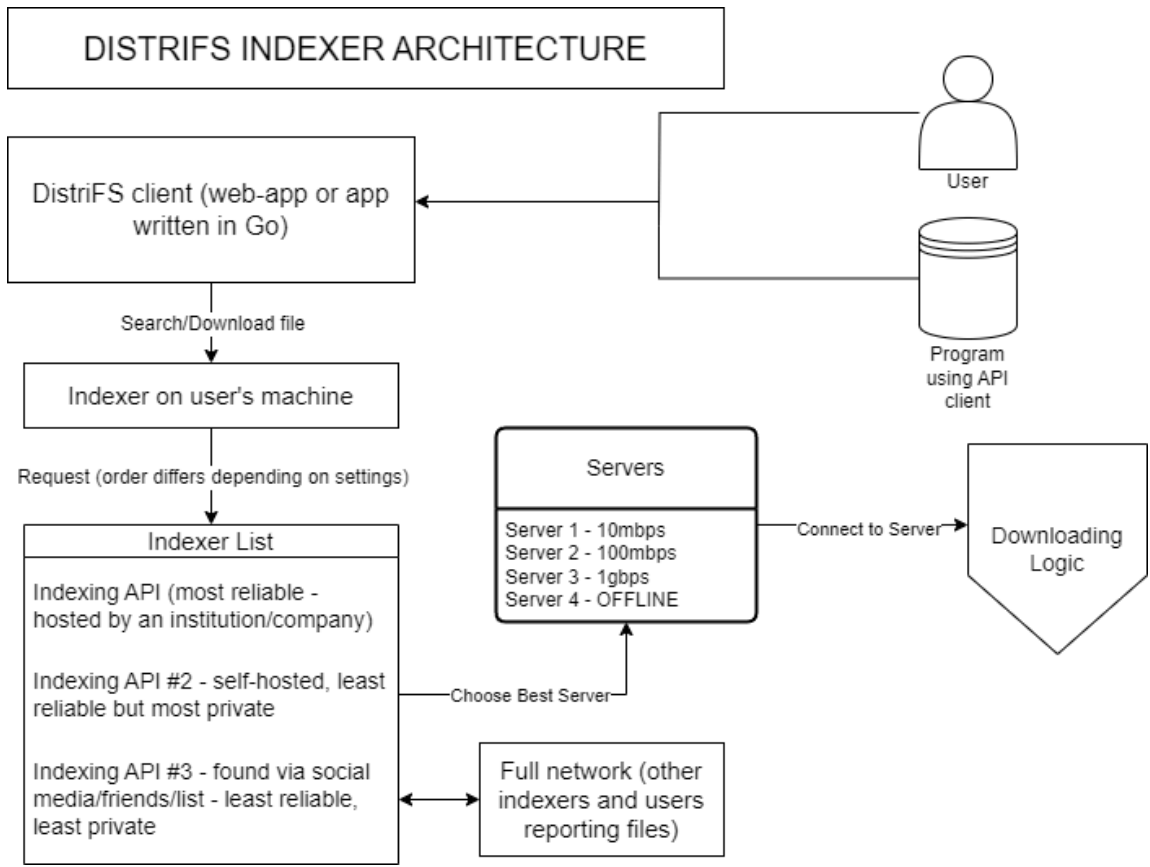

The client, being a key part to a secure and fast filesystem network, is implemented in Go using the philosophy of "sensible defaults". The most important part of the client is the connection to the indexer, as files are searched and verified through the indexer. As shown in Fig. 1, DistriFS clients have access to an indexer, which caches frequently downloaded files and manages connections with other indexers. On the web-based DisriFS client, this "indexer" is simply an interface to connect to third-party hosted indexers[7]. After the user queries a hash or search string, the indexer finds all available servers that host the file. If it is unable to find any data locally, it queries other indexers - default indexers will be automatically added when the client is first started. The user is also able to add and remove indexers as they wish, including the default and official ones. Official indexers should contain the most files on the network, as all other indexers automatically communicate hosted file names and hashes with it. If no file is found on the indexers queried, the user is notified and the download fails - this could be because the user has an incorrect or invalid hash, or no file exists with the name that they searched for. If a file or multiple files are found, a list is returned, with servers and the download speeds (shown in the bolded section of Fig. 1). The client intelligently selects the server geographically closest to them (minimizing ping) and the fastest server. After the most optimal server is picked, the download logic is initialized.

Downloading a file is straightforward - just a token and a file hash is needed. First, the client queries the server with the hash of the file they wish to download. Assuming the file exists, the server responds with the metadata associated with the file. The client can use this metadata to verify that it is downloading the correct file - the structure includes fields such as file size, last modified date and the file hash in SHA256 format. Assuming the user confirms, or the client is on permissive security mode (allowing it to execute more risky operations), the server sends back a one-time download token URL. This URL can be opened in the browser, in a terminal with a tool such as curl or wget, or through the official client.

### 3.4 SERVER ARCHITECTURE

The server is the final component of the system. It simply reads files from the target directory (specified by the client), and indexes them with hashes and metadata. The indexer follows the tree that the server specifies, going from the root directory through each subdirectory to discover all files on the server. For this reason, pagination is not required, but a hard cutoff could be implemented on the indexer side to make sure that it is not overloaded with spam files, which could be provided by a malicious server. The server runs each download on a separate thread, allowing for many concurrent downloads. However, if the server reaches its max throughput regularly, the administrator may choose to limit the amount of concurrent downloads through the use of a queue system. The queue system functions based on "download keys". After the client requests to download the file, the request will hang while it is waiting for other downloads to complete. After the key is provided, the user will have access to download their file without the queue. This mitigates possible DoS (denial of service) and DDoS (distributed denial of service) attacks by forcing the clients to have active connections to the server, causing a large amount of expended bandwidth. While one server may have an extremely large queue, real clients are allowed to route to any server they desire: commonly downloaded and censored files would have more peers to download from. Even large state actors such as the North Korean, Chinese and Russian governments would have to use an enormous amount of resources to restrict access to a single file. After the download is requested, the server provides an HTTP file stream to facilitate the download. The ubiquitous nature of such streams allows for implementations in practically any programming language, and access to the DistriFS network with just a cURL client[8].

---

[7] Due to limitations with the browser's sandbox, it is not possible to run an indexer fully on the web. Unfortunately, the browser does not offer ports to listen to, nor does it allow programs to run 24/7. However, the user is still able to use an interface, which functions as a proxy to the fastest available third-party indexer.

## 4 SECURITY

Security is essential to public trust in the DistriFS network. Solutions such as file integrity checks, an integrated antivirus and secure architecture make it difficult to distribute malware on the network. However, privacy (anonymous downloads and uploads) is still preserved, as the security systems used by the network do not rely on trust to operate.

---

[8] To use a cURL client, the user would have to have relevant documentation about how the network works, as the token logic is not commonly seen in client-facing applications. This documentation will be provided publicly on the GitHub repository for developers who want to integrate with the network.

**Fig. 2 - File integrity check system**

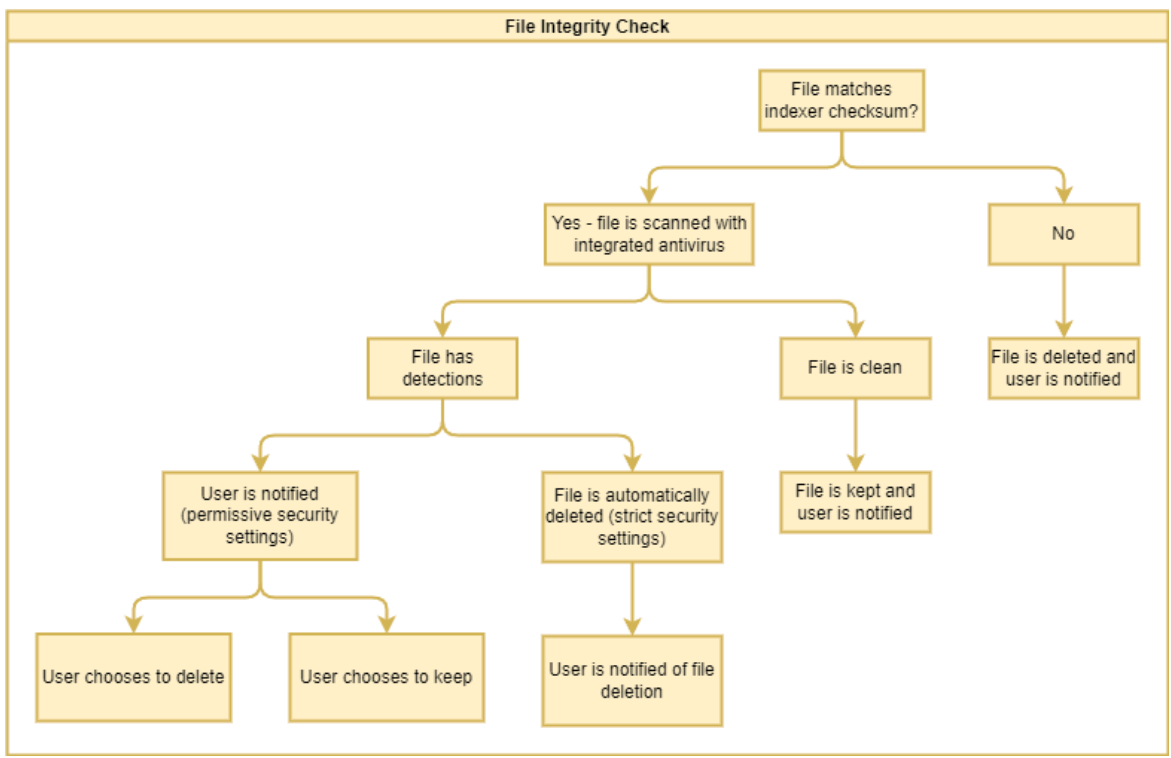

      File integrity checks help keep the user safe from malware and bad actors. If the user knows exactly what they want to download (for example, they are given the hash of a Linux ISO from an official website like ubuntu.com), there is no possibility that DistriFS can be used to distribute malware. After the file is downloaded, the client automatically performs hashing checks to verify that the file downloaded matches the file that the user searched for. Even if both the server and indexer are compromised[9], any unwanted file would be blocked after it is downloaded. As shown in the flowchart (Fig. 2), the file is checked using SHA256 for any discrepancy. However, if the user is using the search function of DistriFS, they incur a higher risk. As anyone can choose any name for a file to host on DistriFS, there is a possibility of malware distribution. As always, the user should stay vigilant when downloading unknown software, and install an antivirus to protect themselves. For additional security, DistriFS includes the open-source antivirus program ClamAV. If the user has enabled virus scanning (which is on by default), they are kept safe from malicious software by ClamAV's ability to detect malware

---

[9] This, while possible, is very unlikely: the client takes a different path each time a file is downloaded by it intelligently selecting the best server

through both static and dynamic detections. For the most cautious users, hashes can be uploaded and scanned with VirusTotal, an integrated third-party service that checks files for malware with 62 different anti-malware scanners. However, this behavior is limited by a 500MB file limit on scans. Additionally, this would only be suitable for files that the user chooses to share with VirusTotal, as the user would not wish to share sensitive information with an online service. Therefore, this scanning is purely opt-in and optional.

## 4.1 PRIVACY

DistriFS upholds a high standard of privacy for the user, as no personally identifying information is required to use the program. However, no system is completely private, and determined organizations and states could still de-anonymize the user if proper precautions are not taken. While the official indexer and server do not take IP (Internet Protocol) address or UA (User-Agent) logs, the source code is public, and may be modified and re-compiled very easily. User-Agents provide a small amount of information on the computer and browser that the file is downloaded on. This risk is completely mitigated through the use of the official DistriFS client, as all headers are changed to protect the user's privacy. The default DistriFS user-agent is "DistriFS/1.0". IPs provide geolocation data on the user, and if the ISP (Internet Service Provider) complies with requests for information from the government, the user may be de-anonymized by police or military actors. This is even more of a concern in countries without rule of law - such records may be obtained instantaneously if the ISP is owned by the state or has close ties to the government. However, as the indexer and server are usually hosted on separate computers, domains and IP addresses, the government would have to own both the indexer and the server to find out the exact file that is downloaded. This provides a layer of privacy, as the client would have to take an unlikely path to encounter both government-operated programs. However, many people, especially those in authoritarian regimes, may not want to take the small risk of being de-anonymized. To mitigate the risk of an IP leak, the Tor Network or a VPN may be used. As VPN companies are usually incorporated in countries which have a very high standard for privacy, a leak of information, even with appropriate legal documents, is very improbable. However, for the most careful of users, the Tor Network may be used. Tor is widely regarded as the gold standard for privacy - however, the privacy that it provides comes at a cost of speed. The Tor Network usually slows download speeds to sub-1 mbps speeds. Therefore, this would be only ideal for the most cautious users.

## 5 CONCLUSION

In this paper, a decentralized, privacy-respecting and fast alternative to other file-distribution solutions is presented. When implemented, this project will significantly advance the field of censorship-resistant file sharing. I hope this project can live up to its expectations as a next-generation, privacy-conscious alternative to current options in file sharing.

**REFERENCES**

Johnson, M. E., McGuire, D., & Willey, N. D. (2008). The evolution of the peer-to-peer file sharing industry and the security risks for users. In *Proceedings of the 41st Annual Hawaii International Conference on System Sciences* (HICSS 2008) (pp. 383-383). Waikoloa, HI, USA. https://doi.org/10.1109/HICSS.2008.436

Burda, D., & Teuteberg, F. (2013). Sustaining accessibility of information through digital preservation: A literature review. *Journal of Information Science*, 39, 442-458. https://doi.org/10.1177/0165551513480107

Morris, P. S. (2009). Pirates of the Internet: At intellectual property's end with torrents and challenges for choice of law. *International Journal of Law and Information Technology*, 17(3), 282-303. https://doi.org/10.1093/ijlit/ean010

Benet, J. (2014). IPFS - Content addressed, versioned, P2P file system. *arXiv preprint* arXiv:1407.3561. https://arxiv.org/abs/1407.3561

Brown, E. (2022). The top reasons countries ask Google to remove content. *ZDNet*. https://www.zdnet.com/article/the-top-reasons-countries-ask-google-to-remove-content/

Thekkath, C. A., Mann, T., & Lee, E. K. (1997). Frangipani: A scalable distributed file system.

*SIGOPS Operating Systems Review*, 31(5), 224–237.

https://doi.org/10.1145/269005.266694

Lu, L., Sankaranarayana Pillai, T., Gopalakrishnan, H., Arpaci-Dusseau, A. C., & Arpaci-Dusseau, R. H. (2017). WiscKey: Separating keys from values in SSD-conscious storage. *ACM Transactions on Storage, 13*(1), Article 5. https://doi.org/10.1145/3033273

Amadeo, R. (2023). Google Drive does a surprise rollout of file limits, locking out some users. *Ars Technica*. https://arstechnica.com/gadgets/2023/03/google-drive-does-a-surprise-rollout-of-file-limits-locking-out-some-users

Cox, R., Griesemer, R., Pike, R., Taylor, I. L., & Thompson, K. (2022). The Go programming language and environment. *Communications of the ACM, 65*(5), 70-78. https://doi.org/10.1145/3488716

Wan, Z., Lo, D., Xia, X., & Cai, L. (2017). Bug characteristics in blockchain systems: A large-scale empirical study. In *Proceedings of the 14th IEEE International Working Conference on Mining Software Repositories: MSR* (pp. 20-21).

Dabbish, L., Stuart, C., Tsay, J., & Herbsleb, J. (2012). Social coding in GitHub: Transparency and collaboration in an open software repository. In *Proceedings of the ACM 2012 Conference on Computer Supported Cooperative Work* (CSCW '12) (pp. 1277-1286). Association for Computing Machinery. https://doi.org/10.1145/2145204.2145396

Munjal, S., Bhardwaj, S., & Malik, S. (2015). Software testing techniques. *In International Journal of Innovative Research in Technology, 1*(11). Retrieved from https://www.ijirt.org/master/publishedpaper/IJIRT101691_PAPER.pdf

Jain, M., Chiu, J., Vardhan, A., Jois, D. & Contributors. (2017). Badger: fast key-value DB in Go. *GitHub Repository*. https://github.com/dgraph-io/badger